\documentstyle[prd,aps]{revtex}

\newcommand{\bea}{\begin{eqnarray}}
\newcommand{\eea}{\end{eqnarray}}

\begin{document}

%%%%%%%%%%%%%%%%%%%%%%%%%%%%%%%%%%%%%%%%%%%%%%%%%%%%%%%%%%%%%%%
\draft
%  For 2 column format.
\twocolumn[\hsize\textwidth\columnwidth\hsize\csname
@twocolumnfalse\endcsname

%%%%%%%%%%%%%%%%%%%%%%%%%%%%%%%%%%%%%%%%%%%%%%%%%%%%%%%%%%%%%%%
\title{{\it COBE}-DMR constraints on a inflation model with 
       non-minimal scalar field}
\author{Jai-chan Hwang${}^{(a)}$ and Hyerim Noh${}^{(b)}$}
\address{${}^{(a)}$ Department of Astronomy and Atmospheric Sciences,
                    Kyungpook National University, Taegu, Korea \\
         ${}^{(b)}$ Korea Astronomy Observatory,
                    San 36-1, Whaam-dong, Yusung-gu, Daejon, Korea \\
         }
\date{\today}
\maketitle

%%%%%%%%%%%%%%%%%%%%%%%%%%%%%%%%%%%%%%%%%%%%%%%%%%%%%%%%%%%%%%%
\begin{abstract}
 
We derive the power spectra of the scalar- and tensor-type structures
generated in a chaotic-type inflation model based on 
non-minimally coupled scalar field with a self interaction.
By comparing contributions of both-structures to the anisotropy
of the cosmic microwave background radiation with the four-year {\it COBE}-DMR
data on quadrupole anisotropy we derive constraints on 
the ratio of self-coupling and non-minimal coupling constants, 
and the expansion rate in the inflation era, Eq. (\ref{constraints}).
The requirement of successful amount of inflation further constrains
the relative amount of tensor-type contribution, Eq. (\ref{r_2-constraint}).

\end{abstract}

\noindent
\pacs{PACS number(s): 98.70.Vc, 98.80.Cq, 98.80.-k}

%%%%%%%%%%%%%%%%%%%%%%%%%%%%%%%%%%%%%%%%%%%%%%%%%%%%%%%%%%%%%%%
%  For 2 column format.
\vskip2pc]
%%%%%%%%%%%%%%%%%%%%%%%%%%%%%%%%%%%%%%%%%%%%%%%%%%%%%%%%%%%%%%%

Immediately after the advent of inflation as a scenario in the early universe,
the inflation is recognized as providing a consistent and successful mechanism 
for the generation and evolution of the observed large-scale cosmic structures,
\cite{inflation}.
In such a scenario, the observed large-scale structures, in return, can
provide constraints on the physics during the inflation era which is
usually based on model scalar fields or generalized versions of gravity
theories.
In particular, the observed anisotropies of the
cosmic microwave background radiation (CMBR) in the large-scale
(being still in linear regime and uncontaminated by later evolution
of small-scale structures) is often regarded as providing a clean window 
to probe the inflation era.
In fact, probing the early universe, thus the 
corresponding high-energy physics, using this observational `window' became
a popular activity in this active field connecting the early universe physics
and the large-scale observation projects.
At the moment, however, the presently available observational data does not
constrain uniquely the early universe physics, and designing the early universe
based on the observation is possible only for certain restricted situations;
for example, designing the field potential of a minimally coupled scalar 
field assuming the slow-roll approximation, \cite{SBB,design-inflation}.
Usually, most approaches are taken in the other way.
That is, if a plausible inflation scenario is made we can use the 
CMBR observation data to constrain the model parameters.
The extreme high-level isotropy of the CMBR provides strong constraints,
and often excludes the considered model from the list of candidate for
a successful inflation.
In this paper, using a recent CMBR data we will derive constraints on 
a candidate based on non-minimally coupled scalar field and suggest
a viable parameter range tolerable by future CMBR experiments.

In \cite{Fakir-Unruh} a chaotic-type inflation scenario was proposed based on
strong-coupling regime of the non-minimally coupled scalar field with a
self-coupling.
As in the ordinary chaotic inflation scenario, the extreme low-level 
anisotropy of the CMBR gives strong constraints on the model parameters.
In the literature, such constraints were derived either using the 
scalar-type structure \cite{SBB,Fakir-Unruh,Makino-Sasaki} or the 
tensor-type structure \cite{Komatsu-Futamase}.
In this paper using both-structures and the four-year {\it COBE}-DMR data
we will derive the proper constraints 
on the coupling constants and Hubble parameter in the inflation era: 
the result is in Eq. (\ref{constraints}).
Condition of enough e-folds during the inflation gives an additional 
constraint on the relative strength of the gravitational wave
compared with the scalar-type structure: see Eq. (\ref{r_2-constraint}).

In \cite{GGT-scalar,GGT-GW} we analysed the quantum generation
and the classical evolution processes of the scalar- and tensor-type
structures based on the following generalized gravity
\bea
   {\cal L} = \sqrt{-g}
       \left[ {1 \over 2} f (\phi, R)
       - {1 \over 2} \omega (\phi) \phi^{;c} \phi_{,c}
       - V (\phi) \right]. 
   \label{Lagrangian}
\eea
Equations for the spatially homogeneous, isotropic and flat background are 
in Eq. (5) of \cite{GGT-scalar};
for a more complete set, see Eqs. (38,51-53) in \cite{GGT-HN}.
The on-shell Lagrangians for the scalar- and tensor-type structures
are presented in Eq. (7) of \cite{GGT-scalar}
and Eq. (3,24) of \cite{GGT-GW}, respectively.
The non-minimally coupled scalar field is a case
with $f = (\kappa^{-2} - \xi \phi^2)R$,
thus $F \equiv f_{,R} = \kappa^{-2} - \xi \phi^2$, and $\omega = 1$ where
$\kappa^2 \equiv 8 \pi m_{pl}^{-2}$.
We consider a self-coupling $V = {1 \over 4} \lambda \phi^4$.
Thus, in the following we will use freely the results 
derived in \cite{GGT-scalar,GGT-GW}.

{\it Assuming} the slow-rolls ($|\ddot \phi /\dot \phi| \ll H \equiv \dot a/a$ 
and $|\dot \phi/\phi| \ll H$), 
the potential-dominance (${1 \over 2} (1 - 6 \xi) \dot \phi^2 \ll V$), 
and the strong-coupling ($|\kappa^2 \xi \phi^2| \gg 1$) conditions
we have the following solution for the background, 
\cite{Fakir-Unruh,Makino-Sasaki,Komatsu-Futamase}: 
\bea
   H = H_i + {\lambda \over 3 \kappa^2 \xi ( 1 - 6 \xi)} 
       \left( t - t_i \right), \quad
       \phi = \sqrt{ - 12 {\xi \over \lambda} } H,
   \label{BG}
\eea
where we consider $\xi < 0$ case \cite{Barvinsky}.
In the regime $H_i$ term {\it dominates} $H$ (we call it a near-exponential
condition) we have
a near exponentially expanding period $a \propto e^{H_i t}$ which can
provide a plausible inflation scenario, \cite{Fakir-Unruh}.

The background evolution in Eq. (\ref{BG}) leads to $n = 2$ in
Eq. (21) of \cite{GGT-scalar} and $n_g = 2$ in Eq. (11) of \cite{GGT-GW}.
Thus, the general mode-function solutions in Eq. (14) of \cite{GGT-scalar}
and Eq. (29) of \cite{GGT-GW} include our cases as special subsets.
In the large-scale limit, the general power spectra 
based on vacuum expectation values in Eq. (16) of \cite{GGT-scalar} 
and Eq. (32) of \cite{GGT-GW} lead to the following:
\bea
   & & {\cal P}_{\hat \varphi_{\delta \phi}}^{1/2} 
       = {H \over |\dot \phi|} {\cal P}_{\delta \hat \phi_\varphi}^{1/2} 
       = {H^2 \over 2 \pi |\dot \phi|} {1 \over \sqrt{1 - 6 \xi} } 
       \big| c_2 (k) - c_1 (k) \big|,
   \label{P-scalar-general} \\
   & & {\cal P}_{\hat C_{\alpha\beta}}^{1/2} 
       = {\kappa H \over \sqrt{2} \pi} {1 \over \sqrt{1 - \kappa^2 \xi \phi^2} }
       \sqrt{ {1 \over 2} \sum_\ell 
       \big| c_{\ell 2} (k) - c_{\ell 1} (k) \big|^2 },
   \label{P-GW-general}
\eea
where $\ell$ indicates the two polarization states of the gravitational wave.
In these forms, $\xi = 0$ reproduces correctly the minimally coupled limit,
\cite{Stewart-Lyth};
see Eqs. (56,6) of \cite{GGT-scalar} and Eq. (34) of \cite{GGT-GW}
[Eq. (\ref{P-GW-general}) is not applicable for general $\kappa^2 \xi \phi^2$,
though]. 
$c_i(k)$ and $c_{\ell i}(k)$ are constrained by the quantization 
conditions: $|c_2|^2 - |c_1|^2 = 1$, and $|c_{\ell 2}|^2 - |c_{\ell 1}|^2 = 1$.
Notice the general dependences of the power spectra on the scale $k$
through the vacuum choices which fix $c_i$ and $c_{\ell i}$.
If we choose the simplest vacuum states $c_2 = 1$ and $c_{\ell 2} = 1$
the power spectra are independent of $k$, thus are scale invariant, 
\cite{comment:scalar-previous}.

Using Eq. (\ref{BG}), Eqs. (\ref{P-scalar-general},\ref{P-GW-general}) become:
\bea
   & & {\cal P}_{\hat \varphi_{\delta \phi}}^{1/2}
       = \left( {H_i \over m_{pl}} \right)^2 
       \sqrt{ - { 12 \xi ( 1 - 6 \xi) \over \lambda} }
       \big| c_2 - c_1 \big|, 
   \label{P-scalar-NM-quantum} \\
   & & {\cal P}_{\hat C_{\alpha\beta}}^{1/2}
       = {1 \over 2 \pi} \sqrt{\lambda \over 6 \xi^2}
       \sqrt{ {1 \over 2} \sum_\ell \big| c_{\ell 2} - c_{\ell 1} \big|^2 }.
   \label{P-GW-NM-quantum} 
\eea
By identifying 
the power spectra based on the vacuum expectation values
during the inflation era [${\cal P}_{\hat \varphi_{\delta \phi}}$ and 
${\cal P}_{\hat C_{\alpha\beta}}$]
with the classical power spectra based on the spatial averages
[${\cal P}_{\varphi_{\delta \phi}}$ and ${\cal P}_{C_{\alpha\beta}}$]
(both in the large-scale limit), we have the same results in 
Eqs. (\ref{P-scalar-NM-quantum},\ref{P-GW-NM-quantum})
now valid for ${\cal P}_{\varphi_{\delta \phi}}$ and 
${\cal P}_{C_{\alpha\beta}}$, \cite{comment:classicalization}.

In \cite{GGT-scalar,GGT-GW} we have shown that, ignoring the transient 
solutions, $\varphi_{\delta \phi}$ and $C_{\alpha\beta}$ are 
{\it conserved} independently of changing gravity, 
changing potential, and changing equation of state, as long as the
scale remains in the large-scale limit; this is the case for the
observationally relevant scales before the second (inward) horizon crossing
in the matter dominated era.
Consequently, even in the matter dominated era 
Eqs. (\ref{P-scalar-NM-quantum},\ref{P-GW-NM-quantum})
remain valid as the power spectra in the large-scale limits 
[the super-sound-horizon for the scalar-type structure 
and the super-horizon for the gravitational wave].
Thus, {\it choosing} the simplest vacuum in the inflation era, 
we have the final classical spectra in the large-scale limit
in matter dominated era as \cite{comment:GW-previous}:
\bea
   & & {\cal P}_{\varphi_{\delta \phi}}^{1/2}
       = \left( {H_i \over m_{pl}} \right)^2 
       \sqrt{ - { 12 \xi ( 1 - 6 \xi) \over \lambda} }, 
   \label{P-scalar} \\
   & & {\cal P}_{C_{\alpha\beta}}^{1/2}
       = {1 \over 2 \pi} \sqrt{\lambda \over 6 \xi^2}. 
   \label{P-GW} 
\eea
{}From these power spectra we can derive the rest of the power spectra
of the classically fluctuating quantities: density, velocity, and
potential fluctuations, and the anisotropy in the CMBR.
{}For example, the rotationally invariant multipole of the anisotropy,
$\langle a_l^2 \rangle \equiv \langle |a_{lm}|^2 \rangle$ 
is related to ${\cal P}_{\varphi_{\delta \phi}}$
and ${\cal P}_{C_{\alpha\beta}}$ through formulae
in Eq. (61) of \cite{GGT-scalar} and Eq. (56) of \cite{GGT-GW}, 
\cite{comment:notation}. 
Thus, the observed values (or limits) of $\langle a_l^2 \rangle$ constrain 
directly ${\cal P}_{\varphi_{\delta \phi}}$ and ${\cal P}_{C_{\alpha\beta}}$, 
thus constrain $H_i$ and $\lambda /\xi^2$.

{}For the scale independent spectra in Eqs. (\ref{P-scalar},\ref{P-GW})
$\langle a_l^2 \rangle$ can be integrated.
The quadrupole anisotropy is the following
\cite{comment:spectrum-derivation}
\bea
   \langle a_2^2 \rangle 
   &=& \langle a_2^2 \rangle_S + \langle a_2^2 \rangle_T
   \nonumber \\
   &=& {\pi \over 75} {\cal P}_{\varphi_{\delta \phi}}
       + 7.74 {1 \over 5} {3 \over 32} {\cal P}_{C_{\alpha\beta}}.
   \label{a_2}
\eea
The four-year {\it COBE}-DMR data shows \cite{COBE}:
\bea
   & & Q_{\rm rms-PS} = 18 \pm 1.6 \mu K, \quad
       T_0 = 2.725 \pm 0.020 K,
   \\
   & & \langle a_2^2 \rangle 
       = {4 \pi \over 5} \left( {Q_{\rm rms-PS} \over T_0} \right)^2 
       \simeq 1.1 \times 10^{-10},
   \label{a_2-value}
\eea
where $Q_{\rm rms-PS}$ is the quadrupole anisotropy normalized to fit the
entire power spectrum for a scale independent spectrum.

Depending on the dominance of the scalar- or tensor-type structures,
from Eqs. (\ref{P-scalar}-\ref{a_2-value})
we can derive constraints on $H_i^2$ and $\lambda/\xi^2$.
In order to alleviate the constraint on $\lambda$ we consider a case 
with $|\xi| \gg 1$.
Introducing a ratio 
$r_2 \equiv \langle a_2^2 \rangle_T / \langle a_2^2 \rangle_S$ we have:
\bea
   & & {\lambda \over \xi^2} = {r_2 \over 1 + r_2} 
       {1280 \pi^2 \over 7.74} \langle a_2^2 \rangle
       \simeq 1.8 \times 10^{-7} {r_2 \over 1 + r_2}, 
   \nonumber \\
   & & {H_i \over m_{pl}} 
%       = \sqrt{ {\sqrt{r_2} \over 1 + r_2}
%       \sqrt{4000 \pi \over 3 \times 7.74} \langle a_2^2 \rangle }
       \simeq 5.1 \times 10^{-5} \sqrt{ {\sqrt{r_2} \over 1 + r_2} }.
   \label{constraints}
\eea
At this point, let us check the conditions used for Eq. (\ref{BG}).
Both the slow-roll and the potential dominance conditions lead to
$|\dot H/H^2| \ll 1$ which gives $r_2 \ll 43$.
{}From Eq. (\ref{BG}) we can show 
$\kappa^2 |\xi| \phi^2 \simeq 4.3/\sqrt{r_2}$.
Thus, the strong-coupling condition requires $r_2 \ll 18$.
The near-exponential condition will be used later, 
see Eq. (\ref{enough-e-fold}).

In ordinary chaotic inflation based on a minimally coupled scalar field
with self-coupling, the observed quadrupole anisotropy severely constrains
$\lambda$ \cite{inflation,design-inflation}, see however \cite{Hu-Morikawa}.
The original motivation for considering the strong non-minimal coupling
is to relax the strong constraint on $\lambda$ by introducing a large $\xi$
\cite{Fakir-Unruh,Makino-Sasaki}: the result is in Eq. (\ref{constraints}).

Authors of \cite{SBB,Fakir-Unruh,Makino-Sasaki} considered only the
scalar-type structure. 
Together with the condition of successful inflation with enough 
number of e-folds $N \sim H_i (t_e - t_b) > 70$ 
[$t_b$ and $t_e$ are the beginning and the ending epochs of inflation] 
they derived a rough constraint on $\lambda/\xi^2$.
The e-folding number is roughly estimated as
\bea
   N \sim \int_{t_b}^{t_e} H dt
             \sim {3 \over 4} \kappa^2 |\xi| \phi^2 \Big|_e^b
             \sim 72 \pi {\xi^2 \over \lambda} 
             \left( {H_i \over m_{pl} } \right)^2. 
   \label{N-def}
\eea
In terms of this estimate of $N$, Eq. (\ref{P-scalar}) becomes
\bea
   {\cal P}_{\varphi_{\delta \phi}}^{1/2}
       \sim {N \over 6 \sqrt{2} \pi } \sqrt{\lambda \over \xi^2},
   \label{P-scalar-N}
\eea
which is $N/\sqrt{3}$ times ${\cal P}_{C_{\alpha\beta}}^{1/2}$ 
in Eq. (\ref{P-GW}).
In such a case, a constraint on $\lambda /\xi^2$ was derived 
in \cite{SBB,Fakir-Unruh,Makino-Sasaki} using $N \sim 70$. 
However, $N$ is not a fixed parameter, and the relation in 
Eq. (\ref{N-def}) almost violates the near-exponential condition.
Authors of \cite{Komatsu-Futamase} considered the gravitational wave, 
but made an error indicated in \cite{comment:GW-previous}.

As a matter of fact, using $N > 70$ we can derive a further strong constraint
on the gravitational wave contribution, $r_2$.
The near-exponential condition on $H$ in Eq. (\ref{BG}) 
implies that, during inflation $t_b < t < t_e$, we have 
$H_i |t-t_i| \ll {3 \over 2} \kappa^2 |\xi| \phi^2$,
and the enough e-folds condition leads to 
\bea
   70 < N \sim 2 H_i (t_e- t_i) \ll 3 \kappa^2 |\xi| \phi^2
       \simeq 13/\sqrt{r_2}.
   \label{enough-e-fold}
\eea
Thus, we have
\bea
   r_2 \ll 0.034,
   \label{r_2-constraint}
\eea
which is a much stronger constraint than from the other conditions
below Eq. (\ref{constraints}).

Similar analyses can be made in the case of the induced gravity
which is a case of Eq. (\ref{Lagrangian}) with
$f = \epsilon \phi^2 R$, $\omega = 1$, and 
$V = {1 \over 4} \lambda (\phi^2 - v^2 )^2$.
A chaotic-type inflation is possible with an assumption $\phi^2 \gg v^2$
\cite{Spokoiny,Fakir-Unruh}; with this condition the inflation scenario
is exactly the same as the one based on a nonminimally-coupled scalar field
with strong coupling and $V = {1 \over 4} \lambda \phi^4$.
By replacing $\xi \rightarrow - \epsilon$ and $\xi \kappa^2 \rightarrow - v^2$,
our analyses and results also apply exactly to the inflation based on
induced gravity; the only difference appears in Eq. (\ref{P-GW-general}) 
because we have written it in a general form including Einstein gravity limit.

In summary, if a non-minimally coupled scalar field with a self-coupling
(or induced gravity) provides a chaotic-type inflation scenario generating 
the observed large-scale structures, the four-year {\it COBE}-DMR data gives 
constraints on the ratio of parameters and the expansion rate in the 
inflation era: see Eq. (\ref{constraints}).
A condition of successful inflation with enough e-folds gives a strong
upper-limit on the gravitational wave contribution: 
see Eq. (\ref{r_2-constraint}).
Therefore, as a conclusion we would like to mention that, 
although the present observation allows only narrow parameter ranges, 
the chaotic-type inflation model based on non-minimally coupled scalar 
field or induced gravity still provides a viable scenario for 
the early universe.
However, any excessive amount of the gravitational wave detected
in future CMBR experiments can rule out the possibility.
Forecasts on detecting the gravitational wave contributions to the
CMBR anisotropies and polarizations in future CMBR experiments using planned
MAP and Planck satellite missions with high-accuracy and small angular
resolution are investigated in \cite{CMBR-forecasts}.

\vskip .5cm
%%%%%%%%%%%%%%%%%%%%%%%%%%%%%%%%%%%%%%%%%%%%%%%%%%%%%%%%%%%%%%%
We thank R. Fakir for drawing our attention to the subject and
M. Sasaki and E. Komatsu for useful suggestions and comments. 
JH was supported by the KOSEF, Grant No. 95-0702-04-01-3 and through the 
SRC program of SNU-CTP.

%%%%%%%%%%%%%%%%%%%%%%%%%%%%%%%%%%%%%%%%%%%%%%%%%%%%%%%%%%%%%%%

%%%%%%%%%%%%%%%%%%%%%%%%%%%%%%%%%%%%%%%%%%%%%%%%%%%%%%%%%%%%%%%

\begin{thebibliography}{99}
\bibitem{comment:scalar-previous}
         Equation (\ref{P-scalar-general}) in the simple vacuum state 
         was derived in \cite{SBB,Fakir-Unruh,Makino-Sasaki} 
         using the conformal transformation method.
         The results in \cite{GGT-scalar} were derived in the original frame of
         generalized gravity; we also showed that the same results can be 
         derived from Einstein frame through conformal transformation.
\bibitem{comment:classicalization}
         We have ignored the possible classicalization factors, 
         $Q(k)$ and $Q_\ell (k)$, 
         introduced in Eq. (59) of \cite{GGT-scalar} and
         Eq. (49) of \cite{GGT-GW}.
\bibitem{comment:GW-previous}
         The gravitational wave spectrum derived in \cite{Komatsu-Futamase} 
         differs from our result.
         Equation (3.35) of \cite{Komatsu-Futamase} is missing
         $(1 + \psi )^{-1}$ factor evaluated in the inflation era. 
         It is equivalent to missing $1/\sqrt{1 - \kappa^2 \xi \phi^2}$ term
         in our Eq. (\ref{P-GW-general}) which is a crucial factor
         in the strong-coupling regime.
         Such a correction will affect the rest of the results 
         in \cite{Komatsu-Futamase}.
         Authors of \cite{Komatsu-Futamase} informed us that an erratum 
         about this point has been prepared, and a new work is in progress
         with results consistent with ours \cite{KF-correction}.
\bibitem{comment:notation}
         In \cite{GGT-scalar,GGT-GW} we used
         ${\cal P}_{\varphi_{\delta \phi}} = {\cal P}_C$ and
         ${\cal P}_{C_{\alpha\beta}} = (2/\pi^2) A_T$ in the large-scale limit.
\bibitem{comment:spectrum-derivation}
         {}For the scalar-type contribution an analytic form integration
         of Eq. (61) in \cite{GGT-scalar} is available.
         {}For the tensor-type contribution we need to integrate 
         Eq. (56) of \cite{GGT-GW} numerically.
         The numerical value can be found in \cite{GW-value} as
         $5 \langle a_2^2 \rangle_T = 7.74 v$;
         in our notation $v = {3 \over 32} {\cal P}_{C_{\alpha\beta}}$.
\end{thebibliography}
\end{document}